\documentclass[a4paper,twocolumn,showpacs]{revtex4-1}
\usepackage[latin1]{inputenc}
\usepackage{amsmath,amssymb}
\usepackage{theorem}
\usepackage{graphicx}
\usepackage{color}
\usepackage{wasysym}

\definecolor{blue}{rgb}{0,0,1}
\definecolor{green}{rgb}{0,1,0}
\definecolor{red}{rgb}{1,0,0}
\definecolor{vio}{rgb}{1,0,1}
\definecolor{ama}{rgb}{1,1,0}
\newcommand{\bc}{\begin{center}}
\newcommand{\ec}{\end{center}}
\newcommand{\be}{\nopagebreak[3]\begin{equation}}
\newcommand{\ee}{\end{equation}}
\newcommand{\ba}{\nopagebreak[3]\begin{eqnarray}}
\newcommand{\ea}{\end{eqnarray}}

\begin{document}

\title{
Explicit expressions for optical scalars in gravitational
lensing from general matter sources}
 
\author{
Emanuel Gallo 
and
Osvaldo M. Moreschi
\\
\vspace{3mm}
\small Facultad de Matem\'atica, Astronom\'{i}a y F\'{i}sica, FaMAF, Universidad Nacional de C\'{o}rdoba\\
\small Instituto de F\'{i}sica Enrique Gaviola, IFEG, CONICET\\
\small Ciudad Universitaria, (5000) C\'{o}rdoba, Argentina. \\
\vspace{3mm}
}

\begin{abstract}

We present explicit expressions for the optical scalars and the deflection angle in terms of the energy-momentum tensor components of matter distributions. Our work generalizes standard references in the literature where normally stringent assumptions are made on the sources.
\end{abstract}

\maketitle


\vspace{5mm}

%
\section{Introduction}
Gravitational lensing has became a significant tool to make progress in our knowledge on the 
matter content of our Universe. 
In particular, there is a large number of works that use gravitational lensing 
techniques in order to know how much mass are in galaxies or clusters of galaxies. 
One of the most exiting results was to reaffirm the need for some kind of
dark matter, that appears to interact with the barionic matter only through gravitation.

The question in which there is yet not general agreement is on 
the very nature of this dark matter.
The most common conception is that it is based on collisionless particles\cite{Weinberg08}, and 
where the pressures are negligible.
However in the context of cosmological studies, one often recurs to models of dark matter
in terms of scalar fields. There is also the possibility that dark matter were 
described in terms of spinor fields.

One method to study the nature of dark matter consists in 
observing the deformation of images of galaxies behind a matter distribution that is the
source of a gravitational lens.

\begin{figure}[h]\label{f:fig0}
\includegraphics[clip,width=0.3\textwidth]{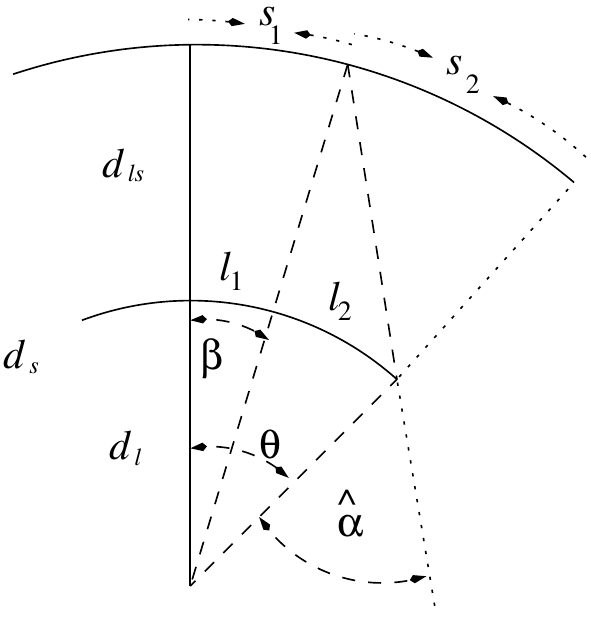}
\caption{This graph shows the basic and familiar angular variables in terms of a simple
flat background geometry.
The letter $s$ denote sources, the letter $l$ denotes lens and the observer is
assumed to be situated at the apex of the rays.}
\end{figure}

In many astrophysical situations, the gravitational effects on light rays is weak, and 
the source and observer are far away from the lens, therefore they are studied under the 
formalism of weak and thin gravitational lenses. The basic and familiar variables in this 
discussion are shown in figure  \ref{f:fig0}.

In this framework the lens equation reads
\begin{equation}
\beta^a=\theta^a-\frac{d_{ls}}{d_s}\alpha^a.
\end{equation}

The differential of this equation can be written as
\begin{equation}\label{eq:standarlensequation}
\delta \beta^a=A^a_{\; b}\,  \delta\theta^b,
\end{equation}
where the matric $A^a_{\; b}$ is in turn expressed by
\begin{equation}
A^a_{\; b} =
\left( {\begin{array}{cc}
 1-\kappa-\gamma_1 & -\gamma_2  \\
 -\gamma_2& 1-\kappa+\gamma_1  \\
 \end{array} } \right);
\end{equation}
where the optical scalars $\kappa$, $\gamma_1$ and $\gamma_2$, are known as 
convergence $\kappa$ and shear components $\{\gamma_1,\gamma_2\}$, and have the information 
of distortion of the image of the source due to the lens effects.
  
It is somehow striking that in
most astronomical works on weak gravitational lensing, it is assumed that the 
lens scalars and deflection angle, can be obtained from a Newtonian-like potential function. 
These expressions although are easy to use, have some limitations:
\begin{itemize}
\item They neglect more general distribution of energy-momentum tensor $T_{ab}$, in particular 
they only take into account the timelike component of this tensor.
In this way they 
severely restrict the possible candidates to dark matter that can be studied with these 
expressions. 
\item They are not expressed in terms of gauge invariant quantities.
\item Since these expressions are written in terms of a potential function, 
it is not easily seen how  different components of $T_{ab}$ contribute in the 
generation of these images.
\end{itemize}

Moreover, most of them assume from the beginning that thin lens is a good approximation.

In other cases in which the thin lens approximation is not used\cite{Bernardeau10}, the results are
presented in a way in which gauge invariance is not obvious.

Here we extend the work appearing in standard references on gravitational 
lensing\cite{Schneider92,Seitz94,Wambsganss98,Bartelmann10} and
present new expressions that do not suffer from 
the limitations mentioned above. In particular we present gauge invariant expressions
for the optical scalars and deflection angle for some general class of matter
distributions.

\section{Integrated expansion and shear}\label{sec:general}
\subsection{General equations: The geodesic deviation equation}
Let us consider the general case of a null geodesic starting from the
position $p_s$ (source) and ending at $p_o$ (observer).
Let us characterize the tangent vector as $\ell= \frac{\partial}{\partial \lambda}$;
so that
\begin{equation}
 \ell^b \nabla_b \ell^a = 0 ;
\end{equation}
that is, $\lambda$ is an affine parameter.

We can now consider also a continuous set of nearby null geodesics.
This congruence of null geodesics can be constructed in the following way.
Let $S$ be a two dimensional spacelike surface (the source image)
such that the null vector $\ell$ is orthogonal to $S$.
Next we can generalize  $\ell$ to be a vector field in the vicinity of the initial geodesic
in the following way: let the function $u$ be defined so that it is
constant along the congruence of null geodesics emanating orthogonally to $S$ and
reaching the observing point  $p_o$.
Then, without loss of generality we can assume that
\begin{equation}
 \ell_a = \nabla_a u ;
\end{equation}
which implies that the congruence has zero twist.

We can complete to a set of null tetrad, so that $m^a$ and $\bar m^a$
are tangent to $S$.
Then a deviation vector at the source image can be expressed by
\begin{equation}
 \varsigma^a = \varsigma \bar m^a + \bar \varsigma m^a .
\end{equation}
In order to propagate this deviation vector along the null congruence
one requires, that its Lie derivate vanishes along the congruence;
that is
\begin{equation}
 \mathfrak{L}_\ell \varsigma^a = 0 .
\end{equation}

Using the GHP notation\cite{Geroch73} it can be shown that the previous equation can be written as
\begin{equation}
\text{\Thorn}(\varsigma)+\zeta\rho+\bar\varsigma\sigma=0;
\end{equation}
where $\text{\Thorn}$ is the well behaved derivation of type $\{1,1\}$ in the direction of $\ell$ (the null geodesic vector of the congruence).

Defining $\mathcal{X}$ by
\begin{equation}
 \mathcal{X} = 
\left(
\begin{array}{c}
\varsigma \\
\bar\varsigma
     \end{array}
\right) ;
\end{equation}
one can prove that $\varsigma$ satisfy
\begin{equation}
 \ell(\ell(\mathcal{X})) = 
 - Q \mathcal{X}
;
\end{equation}
where $Q$ is given by
\begin{equation}
 Q = 
\left(
\begin{array}{cc}
\Phi_{00} & \Psi_{0} \\
\bar\Psi_{0} & \Phi_{00}
     \end{array}
\right) ;
\end{equation}
with
\begin{equation}
 \Phi_{00} = -\frac{1}{2} R_{ab} \ell^a \ell^b ,
\end{equation}
and
\begin{equation}
 \Psi_0 = C_{abcd} \ell^a m^b \ell^c m^d .
\end{equation}
Therefore, this form of the equation only involves curvature quantities.

Although the last equation can be integrated numerically without
problems; it is sometimes convenient to have at hand some
method for approximated solutions.
So, next we present an approximation scheme that it can be applied to any
order one wishes to obtain;
although we will concentrate on the linear approximation since in weak lens studies
it is consistent to consider linear effects of the curvature on geodesic deviations.

Let us first transform to a first order differential equation.
Defining $\mathcal{V}$ to be
\begin{equation}
 \mathcal{V} \equiv \frac{d\mathcal{X}}{d\lambda} ;
\end{equation}
and
\begin{equation}
 \mathbf{X} \equiv 
\left(
\begin{array}{c}
\mathcal{X} \\
\mathcal{V}
     \end{array}
\right) ;
\end{equation}
one obtains
\begin{equation}\label{eq:dotx}
 \ell(\mathbf{X}) = \frac{d\mathbf{X}}{d\lambda}
=
\left(
\begin{array}{c}
\mathcal{V} \\
-Q \,\mathcal{X}
     \end{array}
\right) 
= A \, \mathbf{X}
;
\end{equation}
with
\begin{equation}
 A \equiv
\left(
\begin{array}{cc}
0 & \mathbb{I} \\
-Q & 0
     \end{array}
\right) .
\end{equation}
Equation (\ref{eq:dotx}) can be reexpressed in integral form,
which gives
\begin{equation}
 \mathbf{X}(\lambda) = \mathbf{X}_0 + \int_{\lambda_0}^\lambda A(\lambda') \, \mathbf{X}(\lambda') \, d\lambda' .
\end{equation}
One can define the sequence
\begin{equation}
 \mathbf{X}_1(\lambda) = \mathbf{X}_0 + \int_{\lambda_0}^\lambda A(\lambda') \, \mathbf{X}_0 \, d\lambda'
,
\end{equation}
\begin{equation}
 \mathbf{X}_2(\lambda) = \mathbf{X}_0 + \int_{\lambda_0}^\lambda A(\lambda') \, \mathbf{X}_1(\lambda') \, d\lambda'
;
\end{equation}
and so on.

Assuming that $Q$ is in some sense small, one expects that this sequence will
converge and therefore provide for the solution.

The complete linear iteration is given by
\begin{widetext}
\begin{equation}\label{eq:xlinear}
 \mathbf{X}_3(\lambda) =
\left(
\begin{array}{cc}
\mathbb{I} 
- \int_{\lambda_0}^\lambda 
\int_{\lambda_0}^{\lambda'} 
Q'' \, d\lambda'' \, d\lambda'
& 
(\lambda - \lambda_0) \mathbb{I}
- \int_{\lambda_0}^\lambda 
\int_{\lambda_0}^{\lambda'} 
(\lambda'' - \lambda_0)
Q'' \, d\lambda'' \, d\lambda'
\\
-\int_{\lambda_0}^\lambda  Q' d\lambda'
& 
\mathbb{I}
- \int_{\lambda_0}^\lambda 
(\lambda' - \lambda_0)
Q'  \, d\lambda'
     \end{array}
\right)
\mathbf{X}_0 
;
\end{equation}
\end{widetext}
Now in order to integrate the geodesic deviation equation, we must choose the correct initial conditions.
In the case of light rays belonging to the past null cone of the observer and intersecting 
$S$ at the source,
this initial conditions are
$\mathcal{X} = 0$  and $\mathcal{V} \neq 0$;
since one can think the beam, starts backwards in time from the observer
position, and so initially has vanishing departure, but
with nonzero expansion and shear.

Therefore in the linear approximation one has
\begin{equation}\label{eq:zangular}
\begin{split}
 \mathcal{X}(\lambda) =& 
\left( (\lambda - \lambda_0) \mathbb{I} \right. \\
&- \left. \int_{\lambda_0}^{\lambda}(\lambda-\lambda')(\lambda'-\lambda_0)Q'd\lambda'
\right)
\mathcal{V}(\lambda_0) 
;
\end{split}
\end{equation}
and
\begin{equation}
 \mathcal{V}(\lambda) = 
\left( \mathbb{I} 
- \int_{\lambda_0}^\lambda 
(\lambda' - \lambda_0)
Q'  \, d\lambda'
\right)
\mathcal{V}(\lambda_0) 
.
\end{equation}
In these integrations $\lambda_0$ indicates the position at the observer
and from now on, $\lambda_s$ will indicate the position at the source.

We observe from the first expression, that if the metric were flat ($Q=0$), in order to get a 
deviation vector constructed from $\mathcal{X}_{1}$, defined as
$\mathcal{X}$ evaluated at $\lambda_s = \lambda_0+d_s$, one must choose as initial
condition
\begin{equation}
 \mathcal{V}(\lambda_0) = \frac{1}{(\lambda_s - \lambda_0)}
\mathcal{X}(\lambda_s =\lambda_0 {+} d_s)
=
\frac{1}{d_s}
\mathcal{X}_{1}
.
\end{equation}

But in the case of the presence of a gravitational lens, if an 
observer sees an image of ``size" $\mathcal{X}_o$, 
which means $\mathcal{X}_o \equiv d_s \mathcal{V}_o$ (since actually what is observed is
$\mathcal{V}_o = \mathcal{V}(\lambda_0)$~),
then it should be produced by a source of 
size $\mathcal{X}_s=\mathcal{X}(\lambda_s)$, as described by equation (\ref{eq:zangular}).

We set from now on $\lambda_0=0$ and $\lambda_s=d_s$,
then eq.(\ref{eq:zangular}) reduces to
\begin{equation}\label{eq:zangular2}
 \mathcal{X}_s = 
\left(  \mathbb{I} 
-\frac{1}{d_s} \int_{0}^{d_s} 
\lambda'(d_s-\lambda')
Q' \, d\lambda' \right)
\mathcal{X}_o
.
\end{equation}

\subsection{Optical scalars in terms of curvature}
In order to compare with the standard representation of the lens scalar we note that
the original deviation vector in the source will be given by eq.(\ref{eq:zangular2}), 
i.e.
\begin{equation}\label{eq:zangular3}
 \left(
\begin{array}{c}
\varsigma_{s} \\
\bar\varsigma_{s}
     \end{array}
\right)=
\left(  \mathbb{I} 
-\int_{0}^{d_s} 
\frac{\lambda'(d_s-\lambda')}{d_s}
Q' d\lambda'
\right)
\left(
\begin{array}{c}
\varsigma_{o} \\
\bar\varsigma_{o}
     \end{array}
\right)     ;
\end{equation}
if we make the following decomposition into real and imaginary part, 
\begin{eqnarray}
\varsigma_{o}&=&\varsigma_{oR}+i\varsigma_{oI} ,\\
\varsigma_{s}&=&\varsigma_{sR}+i\varsigma_{sI} ,\\
\Psi_{0}&=&\Psi_{0R}+i\Psi_{0I} ;
\end{eqnarray}
we obtain from eq.(\ref{eq:zangular3}) that
\begin{widetext}
\begin{equation}\label{eq:zetaszetao}
\begin{split}
\varsigma_{sR}=&\left(1-\int_{0}^{d_s} 
\frac{\lambda'(d_s-\lambda')}{d_s}
\left(\Phi_{00}'+\Psi_{0R}'\right)d\lambda'\right)\varsigma_{oR}-\left(\int_{0}^{d_s} 
\frac{\lambda'(d_s-\lambda')}{d_s}
\Psi_{0I}'  d\lambda'
\right) \varsigma_{oI},\\
\varsigma_{sI}=&\left(1-\int_{0}^{d_s} 
\frac{\lambda'(d_s-\lambda')}{d_s}
\left(\Phi_{00}'-\Psi_{0R}'\right) d\lambda'\right)\varsigma_{oI}
-\int_{0}^{d_s} 
\frac{\lambda'(d_s-\lambda')}{d_s}
\Psi_{0I}'d\lambda'
\varsigma_{oR}.
\end{split}
\end{equation}
\end{widetext}
Note also that in principle the integration must be made through the actual geodesic 
followed by a photon in its path from the source to observer. However the last 
expressions are valid only in the limit where the linear approximation is valid.
If one considers a linear perturbation from flat spacetime,  then
the curvature components $\Phi_{00}$ and $\Psi_0$ would be already of linear order.
Then, in the context of weak gravitational lensing,
it is consistent to consider a null geodesic in flat spacetime;
since the actual null geodesic can be thought as 
a null geodesic in flat spacetime plus some corrections of higher orders.

Now, in order to compare with the usual expressions for the lens scalars 
$\kappa, \gamma_1$ and $\gamma_2$, 
let us recall that they are defined via the relation eq.(\ref{eq:standarlensequation});
but since it is a linear relation, one can relate the deviation vectors by the same
matrix, namely
\begin{equation}\label{eq:deltabeta}
\varsigma^i_s=A^i_j\varsigma^j_o;
\end{equation}
where $\{\varsigma^i_s,\,\varsigma^i_o\}$ are the spatial vector associated with
$\{\varsigma_s,\,\varsigma_o\}$ respectively.

Therefore, by replacing into eq.(\ref{eq:deltabeta}), we obtain
\begin{eqnarray}
\varsigma_{sR}&=&(1-\kappa-\gamma_1 )\varsigma_{oR}-\gamma_2\,\varsigma_{oI},\\
\varsigma_{sI}&=&-\gamma_2\,\varsigma_{oR}+(1-\kappa+\gamma_1)\varsigma_{oI};
\end{eqnarray}
which by comparing with eq.(\ref{eq:zetaszetao}), implies that
\begin{eqnarray}\label{eq:lensscalars}
\kappa&=&\frac{1}{d_s}\int_{0}^{d_s} 
\lambda'(d_s-\lambda')
\Phi_{00}'\, d\lambda',\\
\gamma_1&=&\frac{1}{d_s}\int_{0}^{d_s} 
\lambda'(d_s-\lambda')
\Psi_{0R}'\, d\lambda',\\
\gamma_2&=&\frac{1}{d_s}\int_{0}^{d_s} 
\lambda'(d_s-\lambda')\Psi_{0I}'\, d\lambda'.
\end{eqnarray}
Let us emphasize that these expressions for the weak lens quantities are explicitly
gauge invariant, since they are given in terms of the curvature components. This is in contrast to the usual treatment of weak lensing
found in the literature, which use for example equation (2.17) of reference
\cite{Schneider92} as the source for the calculation of the lens scalars.

Note that these expressions are valid for any weak field gravitational lens on a
perturbed at spacetime, without restriction on the size of the lens compared with
the other distances. As a final comment to this section it is important to remark that these expressions are 
valid for any weak gravitational lensing on a perturbed flat spacetime, without 
restriction on the size of the lens compared with the other distances.
However, if we make use of the hypothesis of thin lens, these equations 
can be further simplified, as we will show below. 

\section{The thin lens approximation}\label{sec:thin}
\subsection{The general case}
Now, we will consider the case of a lens whose size is small compared with the distances to 
the source and the observer.
Let us consider a null geodesic coming from a source located 
at a distance $d_s$ from the observer, and at a distance $d_{ls}$ from the lens, 
coming parallel to the $y$ axis, but in the negative direction;
we will use $J$ to represent the impact parameter and  $\vartheta$
to denote the angle of the trajectory as measured from the $z$ axis, in the $(z,x)$ plane.

Then if we assume a thin lens, $\Phi_{00}$ and $\Psi_0$ will be sharply peaked around $\lambda=d_l$, 
where it is located  and the expressions for the lens 
scalars are reduced to
\begin{equation}\label{eq:g1masig}
\begin{split}
\kappa = \frac{d_ld_{ls}}{d_s}\hat{\Phi}_{00},
\end{split}
\end{equation}
\begin{equation}
\begin{split}\label{eq:g1masig2}
\gamma_1+i\gamma_2 = \frac{d_ld_{ls}}{d_s}\hat{{\Psi}}_0,
\end{split}
\end{equation}
where
\begin{equation}
\begin{split}
\hat{\Phi}_{00}=&\int^{d_s}_0\Phi_{00}d\lambda,\\
\hat{\Psi}_{0}=&\int^{d_s}_0\Psi_{0}d\lambda,
\end{split}
\end{equation}
are the projected curvature scalars along the line of sight.

We again emphasize that these expressions for the weak lens scalars are explicitly 
gauge invariants.

\subsection{The axially symmetric case }
\subsubsection{The lens scalars  in terms of projected Ricci and Weyl Scalars}
For axially symmetric lens , 
the projected curvature scalars are given by
\begin{eqnarray}
\hat\Phi_{00}(J)&=&\int^{d_s}_0\Phi_{00}(\lambda')d\lambda',\\
\hat{\Psi}_{0}(J)&=&-e^{2i\vartheta}\hat{\psi}_{0}(J).
\end{eqnarray}
where one can see that
\begin{eqnarray}
 \hat{\psi}_{0}(J) &=&-e^{-2i\vartheta}\int^{d_s}_{0} \Psi_0(\lambda')d\lambda'.
 \end{eqnarray}
The reason for the minus sign choice is that in many common astrophysical situations
one would find $\hat{\psi}_{0}(J) > 0$.

By replacing in eqs. (\ref{eq:g1masig}) and (\ref{eq:g1masig2}), we obtain for the lens scalars
\begin{eqnarray}\label{eq:lensscalarthin}
\kappa &=&\frac{d_{ls} d_l}{d_s}\hat{\Phi}_{00}(J),\\
\gamma_1 &=& - \frac{d_{ls} d_l}{d_s}\hat{\psi}_{0}(J)\cos(2\vartheta),\label{eq:lensscalarthin1}\\
\gamma_2 &=& - \frac{d_{ls} d_l}{d_s}\hat{\psi}_{0}(J)\sin(2\vartheta) ,\label{eq:lensscalarthin2}
\end{eqnarray}
which implies that
\begin{equation}\label{eq:lensscalarthin-g}
\gamma =  \frac{d_{ls} d_l}{d_s}\hat{\psi}_{0}(J) .
\end{equation}
These equations can be compared to those of reference \cite{Frittelli00};
where they use different notation but similar content.

\subsubsection{Deflection angle in terms of projected Ricci and Weyl Scalars}
We wish now to express the deflection angle in terms of the curvature scalars.

If we define the components of $\alpha^i=(\alpha^1,\alpha^2)$ as
\begin{equation}
(\alpha^i)= \alpha(J) (\frac{z_0}{J}, \frac{x_0}{J} );
\end{equation}
then, it can be shown that the optical scalars can be written in terms of $\alpha(J)$ as
\begin{eqnarray}\label{eq:scalarlensalpha}
\kappa &=&\frac{1}{2}\frac{d_{ls}d_l}{d_s}
\left( \frac{d\alpha}{dJ} +  \frac{\alpha(J)}{J}  \right),\\
\gamma_1 &=& \frac{1}{2}\frac{d_{ls}d_l}{d_s} \label{eq:scalarlensalpha1}
\,\cos(2\vartheta) \, \left( \frac{d\alpha}{dJ} -  \frac{\alpha(J)}{J}  \right),\\
\gamma_2 &=&\frac{1}{2}\frac{d_{ls}d_l}{d_s} \label{eq:scalarlensalpha2}
\,\sin(2\vartheta) \, \left( \frac{d\alpha}{dJ} -  \frac{\alpha(J)}{J}  \right).
\end{eqnarray}

It is interesting to note that
\begin{equation}
\kappa-
\gamma_1 \cos(2\vartheta) - \gamma_2 \sin(2\vartheta) 
=\frac{d_ld_{ls}}{d_{s}}\frac{\alpha(J)}{J};
\end{equation}
from which, using  eqs. (\ref{eq:lensscalarthin}-\ref{eq:lensscalarthin2}), it is deduced that
\begin{equation}\label{eq:simplealpha}
\alpha(J)=
J(\hat{\Phi}_{00}(J)+\hat{\psi}_{0}(J)).
\end{equation}
It is worthwhile to remark that this constitutes an equation for the bending angle
expressed in terms of the gauge invariant curvature components in a very simple
compact form.

\section{Detailed study of stationary spherically symmetric lenses }\label{sec:spher}

A stationary spherically symmetric spacetime can be expressed in terms of the standard line element 
\begin{equation}\label{eq:ds1}
 ds^2 = a(r) dt^2 - b(r) dr^2 - r^2(d\theta^2 + \sin^2 \theta d\varphi^2) .
\end{equation}

where it is convenient to define $\Phi(r)$ and $m(r)$ from
\begin{equation}
 a(r) = e^{2 \Phi(r)} ,
\end{equation}
and
\begin{equation}
b(r) = \frac{1}{1 - \frac{2 m(r)}{r} }
.
\end{equation}

For our purpose, it is more convenient to use a null coordinate system to describe the spherically symmetric geometry.
Let us introduce then, a function
\begin{equation}
 u = t - r^* ;
\end{equation}
where $r^*$ is chosen so that $u$ is null. Then by inspection
of equation (\ref{eq:ds1}) one can see that
\begin{equation}\label{eq:du}
 du = dt - \frac{dr^*}{dr} dr = dt - \sqrt{\frac{b}{a}} dr ;
\end{equation}
since then one has
\begin{equation}\label{eq:ds2}
 ds^2 = a\, du^2 + 2 \sqrt{a b} du dr 
        - r^2(d\theta^2 + \sin^2 \theta d\varphi^2) .
\end{equation}

It is natural to define the principal null direction $\tilde\ell_P$
from
\begin{equation}\label{eq:ell}
 \tilde\ell_P = du ;
\end{equation}
which implies that the vector is
\begin{equation}
 \tilde\ell_P^a = g^{ab} du_b = \frac{1}{\sqrt{a b}} 
\left( \frac{\partial}{\partial r}\right)^a .
\end{equation}
We complete to a null tetrad with
\begin{equation}
 \tilde{n}_P = \frac{\partial}{\partial u} + U \mathbb{A} \frac{\partial}{\partial r} ,
\end{equation}
with the complex null vector
\begin{equation}
 \tilde{m}_P = \frac{\sqrt{2} P_0}{r}  \frac{\partial}{\partial \zeta} ;
\end{equation}
in terms of the stereographic coordinate $\zeta$.

Therefore, one has
\begin{equation}
 \mathbb{A} = \frac{1}{\sqrt{a b}} ,
\end{equation}
and
\begin{equation}
 U = - \frac{1}{2 b \mathbb{A}^2} = -\frac{a}{2}
.
\end{equation}

The more general distribution of energy-momentum compatible with spherical 
symmetry is  given by
\begin{equation}
 T_{tt} =  \varrho e^{2 \Phi(r)};
\end{equation}
\begin{equation}
 T_{rr} =    \frac{P_r}{\left( 1 - \frac{2 m(r)}{r} \right) } ;
\end{equation}
\begin{equation}
 T_{\theta \theta} =    P_t \, r^2 ;
\end{equation}
\begin{equation}
 T_{\varphi \varphi} =    P_t \,  r^2 \sin(\theta)^2 ;
\end{equation}
where we have introduced the notion of radial component  $P_r$ and
tangential component  $P_t$.

The Einstein field equations
\begin{equation}
G_{ab} = -8\pi T_{ab} ,
\end{equation} 
in terms of the previous variables are
\begin{equation}\label{eq:rho}
 \frac{dM}{dr} = 4\pi r^2 \varrho ,
\end{equation}
\begin{equation}\label{eq:mgrt}
r^2\frac{d\Phi}{dr} = \frac{m + 4\pi r^3 P_r}{1 - \frac{2 M(r)}{r} } ,
\end{equation}
\begin{equation}
\begin{split}\label{eq:pt}
 r^3 &\left(\frac{d^2\Phi}{dr^2}+(\frac{d\Phi}{dr})^2\right)( 1-\frac{2 M}{r})\\
&+ r^2 \frac{d\Phi}{dr} (1 - \frac{M}{r}-\frac{dM}{dr})\\
&- r \frac{dM}{dr} + M 
= 8\pi r^3 P_t  .
\end{split}
\end{equation}

One can show that the Ricci scalars defined with respect to the tetrad $\{\tilde\ell_P,\tilde n_P,\tilde m_P,\bar{\tilde{m}}_P \}$ are given by
\begin{eqnarray}\label{eq:phis}
 \tilde{\Phi}_{00} &=&\frac{4 \pi}{ a}\left(  
 \varrho + P_r \right),\\
\tilde{\Phi}_{11} &=& \pi\left(  \varrho - P_r + 2  P_t \right),\label{eq:phis11}\\
 \tilde{\Phi}_{22} &=& a \pi \left(  \varrho + P_r \right),\\
\tilde{\Lambda} &=& \frac{\pi}{3}(\varrho -  P_r - 2 P_t);
\end{eqnarray}
and the Weyl scalar,
\begin{equation}\label{eq:psi2rhoP}
\begin{split}
 \tilde{\Psi}_2  
=& \frac{4 \pi}{3} (\varrho - P_r  + P_t)-\frac{m}{r^3} .
\end{split}
\end{equation}
Now, when one made a tetrad transformation between the spherically symmetric tetrad and
the tetrad adapted to the photon geodesic, one find that the function $\alpha(J)$ expressed in
terms of the spherically symmetric null tetrad reads,
\begin{equation}\label{eq:lense-ext2}
\begin{split}
 \alpha(J)
= J \int_{-d_l}^{d_{ls}}  
 &\left[- \frac{3 J^2}{r^2}\tilde{\Psi}_2+\frac{2 J^2}{r^2}(\tilde{\Phi}_{11}
- \frac{1}{4}\tilde\Phi_{00}) \right. \\
&\; + \left. \tilde\Phi_{00}  \right] dy 
.
\end{split}
\end{equation}
Note that in this case, the integration is on the coordinate $y$, instead of
using arbitrary affine parameter. Also note that $r=\sqrt{J^2 + y^2}$.

This constitutes an important explicit relation for the bending angle
in terms of the curvature as seen in an spherically symmetric frame;
which is the natural frame for the sources of the gravitational lens.

In terms of the physical energy-momentum tensor it is obtained
\begin{equation}\label{eq:lense-ext2-b}
\begin{split}
\alpha(J) = J \int_{-d_l}^{d_{ls}} 
 &\left[
 \frac{3 J^2}{r^2} \left( \frac{m(r)}{r^3} -  \frac{4 \pi}{3} \varrho(r) \right)
\right. \\
&\; \left. + 4 \pi \left(\varrho(r) +  P_r(r) \right)
\right] dy 
.\end{split}
\end{equation}

In a similar way, the lens scalars, in terms of the spherically symmetric physical
fields, are given by
\begin{equation}
\begin{split}
\kappa=&\frac{4\pi d_ld_{ls}}{d_s}
\int_{-d_l}^{d_{ls}} 
\left [\rho + P_r + \frac{J^2}{r^2}(P_t-P_r) \right]dy \\
\gamma = & \frac{ d_ld_{ls}}{d_s}
\int_{-d_l}^{d_{ls}} 
\frac{J^2}{r^2}\left [4\pi(\rho+P_t-P_r)-\frac{3 m}{r^3}\right]dy
.
\end{split}
\end{equation}
These new expressions let us see explicitly the contributions of different components of the
energy-momentum tensor on the optical scalars.
Let us note that a couple of terms disappear in the isotropic case in which $P_r=P_t$.

\section{Final comments}\label{sec:final}

Several works on weak gravitational lensing reach up to the expressions that relate the optical scalars
with the curvature components in terms of the tetrad adapted to the motion of the photons;
we have here also presented expressions for the bending angle in terms of the curvature components.
Furthermore, we have presented above expressions for the optical scalars and deflection angle directly in terms
of the matter components of the sources of the gravitational lens, valid for an extended class
of matter distributions.
Our expressions circumvent several deficiencies as are: gauge dependence, lack of explicit expressions,
neglect of spacelike components of the energy-momentum tensor, etc.
The extension of this study to sources with different structure and to the cosmological background 
will be presented elsewhere.

\section*{Acknowledgments}

We acknowledge support from CONICET and SeCyT-UNC.


\end{document}